\documentclass[letter, lineno]{aa}  

\usepackage{graphicx}
\usepackage{txfonts}
\usepackage{xcolor}
\usepackage{lipsum}
\usepackage{subcaption}         % necessary for continued figures, example in section 3
\usepackage{lscape}             % to rotate a single page table, example in appendix.
\usepackage{placeins}           % useful with \FloatBarrier, to keep 
\usepackage{hyperref}    
\graphicspath{ {Figures/} }
\defcitealias{2024MNRAS.531.4275B}{B24}
\defcitealias{2024A&A...688A..15D}{D24}
                                
\begin{document}

%%%%%%%%%%%%%%%%%%%%%%%%%%%%%%%%%%%%%%%%
% if you use custom commands in your title,
% ensure to check your title when submitting!
%%%%%%%%%%%%%%%%%%%%%%%%%%%%%%%%%%%%%%%%
   \title{The 35-Myr old infant planet TOI-837 b has a mildly misaligned orbit }

   %\subtitle{Subtitle}

%%%%%%%%%%%%%%%%%%%%%%%%%%%%%%%%%%%%%%%%
% Please separate each author with the \and command
%
% Please do not include ORCIDs next to author names.
% Only ORCIDs authenticated by individual authors in EDPS
% editorial system will be taken into account.
% ORCIDs included here will be removed.
%%%%%%%%%%%%%%%%%%%%%%%%%%%%%%%%%%%%%%%%

   \author{G. Mantovan
          \inst{\ref{inst1},\ref{inst2}}$^{\orcid{0000-0002-6871-6131}}$, L. Malavolta\inst{\ref{inst2},\ref{inst3}}$^{\orcid{0000-0002-6492-2085}}$, F. Marzari\inst{\ref{inst4}}, A. F. Lanza\inst{\ref{inst5}}$^{\orcid{0000-0001-5928-7251}}$, F. Borsa\inst{\ref{inst6}}$^{\orcid{0000-0003-4830-0590}}$, D. Nardiello\inst{\ref{inst3},\ref{inst2}}$^{\orcid{0000-0003-1149-3659}}$, S. Benatti\inst{\ref{inst7}}$^{\orcid{0000-0002-4638-3495}}$, M. Damasso\inst{\ref{inst8}}$^{\orcid{0000-0001-9984-4278}}$, S. Desidera\inst{\ref{inst2}}$^{\orcid{0000-0001-8613-2589}}$
        }

   \institute{Centro di Ateneo di Studi e Attivit\`a Spaziali ``G. Colombo'' -- Universit\`a degli Studi di Padova, Via Venezia 15, IT-35131, Padova, Italy\label{inst1};
\email{giacomo.mantovan@unipd.it}
            \and
             INAF -- Osservatorio Astronomico di Padova, Vicolo dell'Osservatorio 5, IT-35122, Padova, Italy\label{inst2}
             \and
             Dipartimento di Fisica e Astronomia ``Galileo Galilei'', Università di Padova, Vicolo dell'Osservatorio 3, IT-35122, Padova, Italy\label{inst3}
             \and
             Dipartimento di Fisica e Astronomia, Università di Padova, Via Marzolo 8, IT-35121, Padova,Italy\label{inst4}
             \and
             INAF -- Osservatorio Astrofisico di Catania, Via Santa Sofia 78, IT-95123, Catania, Italy\label{inst5}
             \and
             INAF -- Osservatorio Astronomico di Brera, Via E. Bianchi 46, 23807 Merate (LC), Italy\label{inst6} 
             \and
             INAF -- Osservatorio Astronomico di Palermo, Piazza del Parlamento 1, IT-90134, Palermo, Italy\label{inst7}
             \and 
             INAF -- Osservatorio Astrofisico di Torino, Via Osservatorio 20, IT-10025, Pino Torinese, Italy\label{inst8}
             }

   \date{Compiled: \today}
 
  \abstract 
   {The measurement of the spin-orbit obliquity, that is, the angle between the orbital axis of a planet and the stellar spin axis, provides crucial insights into how planets form and migrate. Observations of young transiting planets, which have not yet experienced significant tidal alterations, offer a unique opportunity to study their original obliquity configuration. We observed the warm Saturn-sized TOI-837~b (member of the 35 Myr old open cluster IC 2602) in-transit using ESPRESSO at VLT, collecting high-precision radial velocities to measure the Rossiter-McLaughlin effect. We found a sky-projected obliquity of $\lambda = 341.1 ^{+2.3}_{-2.5}$ deg. Using our knowledge of the stellar rotation period ($3.00 \pm 0.02$~d), we estimated a true obliquity of $\psi = 25.9 ^{+7.5}_{-6.3}$ deg, which indicates prograde motion and suggests a mildly misaligned orbit. This places TOI-837~b as the first planet younger than 100 Myr with accessible $\psi$ incompatible with an aligned orbit. Together with the primordial circular orbit of TOI-837 b and the presence of a bound stellar companion, this mild misalignment favours the possibility of a primordial obliquity excitation (secular torque on the protoplanetary disc) followed by disc-driven migration, rather than high-eccentricity migration after formation.}

    \authorrunning{Giacomo Mantovan et al.}

   \maketitle

%%%%%%%%%%%%%%%%%%%%%%%%%%%%%%%%%%%%%%%%%%%%%%%%%%%%%%%%%%%%%%
\section{Introduction}

The spin-orbit obliquity is a key diagnostic for the mechanisms of formation and orbital migration of exoplanets \citep[e.g.][]{2011Natur.473..187N}. The sky-projected obliquity ($\lambda$) can be measured with in-transit radial velocities (RV) through the Rossiter-McLaughlin (RM) effect \citep[e.g.][]{2000A&A...359L..13Q, 2005ApJ...622.1118O}. This can then be translated into the 3D orbital obliquity ($\psi$) with respect to the stellar rotation axis with knowledge of stellar rotation period, projected rotational velocity, and stellar radius, yielding the stellar inclination $i_\star$ \citep[e.g.,][]{2007AJ....133.1828W}.

Giant planets in close-in orbits are thought to form in situ close to the final orbit, or in the outer regions and migrate inward \citep{2018ARA&A..56..175D}. Different mechanisms could shrink the orbits, such as dynamical interactions (high eccentricity migration, HEM) through planet-planet scattering \citep{2006A&A...453..341M} or the Kozai mechanism \citep{2003ApJ...589..605W} after disc dispersal, and disc--planet interactions \citep{1996Natur.380..606L}. These mechanisms should imprint distinctive signatures in the observed obliquities. While scattering encounters or the HEM mechanism typically result in misaligned and eccentric orbits, migration through disc-planet interactions (disc-migration) typically preserves low eccentricities and orbital coplanarity. However, star-planet tidal interactions may alter the obliquity, and close-in planets orbiting a few Gyr old stars with deep convective envelopes are expected to show aligned configurations regardless of the formation scenario, as observed in most cases \citep[e.g.][]{2012ApJ...757...18A, Lai12}. By observing systems that are still young enough to not have undergone significant tidal alterations of obliquity, we can access the original configuration in which the system formed \citep[e.g.][]{2022PASP..134h2001A}. Hence, young transiting planets represent a unique resource for solving yet inconclusive observational results \citep[e.g.][]{2023AJ....166..217W, 2024A&A...684L..17M}. 

Space-based transit missions led to the identification of several planets transiting young stars (age $<$ 1 Gyr), totalling about 100 confirmed or validated systems \citep[cf. NASA Exoplanet Archive,][]{2025PSJ.....6..186C}. However, the orbital obliquity has only been measured for 17 young systems: Kepler-63~b \citep{2013ApJ...775...54S}, AU Mic~b \citep{2020A&A...643A..25P}, K2-25~b \citep{2020MNRAS.498L.119G}, DS Tuc A~b \citep{2021A&A...650A..66B}, TOI-942~b \citep{2021ApJ...917L..34W}), V1298 Tau~b and~c \citep{2021AJ....162..213F,2022AJ....163..247J}, HD 63433~b and~c \citep{2020AJ....160..179M,2020AJ....160..193D}, HIP 67522~b \citep{2021ApJ...922L...1H}, TOI-1268~b \citep{2022ApJ...926L...7D}, TOI-1136~d \citep{2023AJ....165...33D}, TOI-2076~b \citep{2023ApJ...944L..41F}, TOI-5398~b \citep{2024A&A...684L..17M, 2024AJ....168..116R}, K2-33~b \citep{2024MNRAS.530.3117H}, TOI-5027~b \citep{2025AJ....170...70E}, Qatar-4~b \citep{2025A&A...694A..91Z}, TOI-2046~b \citep{2025A&A...694A..91Z}, and IRAS 04125+2902~b \citep{2025ApJ...994L..55B}. These measurements have generally indicated aligned configurations, although with moderately large uncertainties. Consequently, the obliquity of young transiting planets remains significantly unexplored. Moreover, among the very small sample of stars with ages $\leq$100 Myr (infant planets), each obliquity measurement is consistent with good alignment \citep{2021ApJ...922L...1H, 2022PASP..134h2001A, 2025ApJ...994L..55B}. It is crucial to extend the sample of such very young stars with obliquity measurements, as this trend would be an important clue about obliquity excitation for giant planets in close orbits. For instance, very young close-orbit giant planets could have formed in situ or undergone disc-migration (both would preserve the initial alignment), while late-arriving ones could have formed through dynamical interactions \citep{2022PASP..134h2001A}.

TOI-837 b \citep{2020AJ....160..239B} is a giant planet in the 35-Myr-old southern open cluster IC 2602 \citep[][B24, D24]{2024MNRAS.531.4275B, 2024A&A...688A..15D}. Open clusters offer the unique opportunity to study exoplanets hosted by cluster members in extreme detail. It is possible to derive properties like $R_\star, M_\star, T_\textrm{eff}, \log \textrm{g}$, distance, age and metallicity with a high level of accuracy. Moreover, the analysis of exoplanets orbiting stars in different open clusters with different ages allows us to trace the temporal evolution of exoplanets, from their early formation stages to their old ages.
%%%%%%%%%%%%%%%%%%%%%%%%%%%%%%%%%%%%%%%%%%%%%%%%%%%%%%%%%%%%%%

\section{Target selection and properties}
TOI-837~b was chosen from the pull of targets selected for the ongoing project ``Obliquity of tidally--detached planets via the Rossiter-McLaughlin effect'' (ORME, Mantovan et al. in prep). The ORME project -- the first result of which was presented in \citet{2024A&A...684L..17M} -- aims to determine the orbital obliquity of young (age $<$ 1 Gyr) or tidally--detached \citep[e.g.][]{2021AJ....162..182R} exoplanets by collecting in-transit RVs to measure the RM effect.

While both confirming studies \citepalias{2024MNRAS.531.4275B, 2024A&A...688A..15D} found that TOI-837~b has a mass of $\approx120 M_\oplus$, they estimated slightly different radii ($R_{\rm p}=9.2^{+0.4}_{-0.3}~R_\oplus$; $R_{\rm p}=9.7^{+0.9}_{-0.6}~R_\oplus$) and densities ($\rho_{\rm p}=0.9 \pm 0.2$~g cm$^{-3}$; $\rho_{\rm p}=0.7 \pm 0.2$~g cm$^{-3}$). They found solutions that are consistent with TOI-837~b having a nearly circular orbit, with an $1\sigma$ upper limit on the eccentricity of $e < 0.27$ \citepalias{2024MNRAS.531.4275B}. The discrepancy in $R_{\rm p}$ measurements is likely due to different dilution treatments in crowded environments, with the work by \citetalias{2024A&A...688A..15D} potentially affected by overcorrection of stellar contamination. Our reanalysis of \textit{TESS} light curves presented in Sect. \ref{sec:analysis} reconciles this discrepancy.

\section{Observations and data reduction}
\subsection{ESPRESSO observations}
We observed TOI-837 with ESPRESSO \citep{2021A&A...645A..96P} 1-UT mode at ESO's Very Large Telescope (VLT) on March 29, 2025 (proposal P114.27GY, PI: G. Mantovan), obtaining 53 spectra with 300 s of exposure time, with an average seeing of 0.4 arcseconds, an $\langle S/N\rangle_{5500 \AA} = 46$ and $\langle {\rm RV}_{\rm err}\rangle = 7.4$ m\,s$^{-1}$. We collected in-transit (2 hours) and suitable out-of-transit (3 hours) observations. The wavelength range of these spectra is 378-789 nm with a resolving power of R $\sim$ 140~000.

We used the ESO's official ESPRESSO pipeline through the EsoReflex workflow \citep{2013A&A...559A..96F}, and computed the RVs using the cross-correlation function (CCF) method \citep{2002Msngr.110....9P}. The strong stellar activity of TOI-837 distorts the core of the average line profile, while its fast stellar rotation broadens the line. Hence, we selected a half-window large enough to include the continuum when fitting the CCF profile. Specifically, we used an F9 mask and a grid size of 120 km s$^{-1}$. 

\subsection{TESS photometry}
\label{sec:tess}
\textit{TESS} observed TOI-837 at 2 min cadence in sectors 10, 11, 37, 38, 63, 64, and 90. We extracted lightcurves using the  \texttt{PATHOS} approach \citep[][]{2020MNRAS.495.4924N}, which allowed us to minimise neighbour flux contamination (down to 0.5 \textit{TESS} pixels) and preserve stellar rotation information more effectively \citep[e.g.][]{2022A&A...664A.163N, 2023A&A...672A.126D, 2024A&A...682A.129M}. Sector 90 coincided with the ESPRESSO observing window, thus enabling a simultaneous observation crucial to better model stellar activity by combining photometric and spectroscopic data.

\section{Analysis}
\label{sec:analysis}
To refine the planetary and stellar parameters reported in \citet{2020AJ....160..239B}, \citetalias{2024MNRAS.531.4275B} and \citetalias{2024A&A...688A..15D}, we examined our light curves and spectroscopic time series within a Bayesian framework using \texttt{PyORBIT}\footnote{\url{https://github.com/LucaMalavolta/PyORBIT}} \citep{2016A&A...588A.118M, 2018AJ....155..107M}, a public software that allows planetary transits and RVs to be modelled while considering stellar activity effects. We also did a transmission spectroscopy analysis to probe the atmospheric composition of TOI-837~b; the results and methodology are summarised in App. \ref{app:atmo}.

\subsection{Joint Bayesian analysis}
We simultaneously modelled all \textit{TESS} light curves and the in-transit RV anomaly. In this way, we modelled the orbital obliquity, the planetary transits, and the stellar activity using Gaussian Processes \citep[GPs,][]{rasmussen2006gaussian, 2014MNRAS.443.2517H, 2015MNRAS.452.2269R}. We fitted the central time of transit ($T_{0,~b}$), orbital period ($P_b$), impact parameter ($b$), planet-to-star radius ratio ($R_{\rm p}/R_\star$), and sky-projected obliquity ($\lambda$). Following \cite{2022MNRAS.516.4432M}, we determined the dilution factor ($df$) caused by stars within 0.5 pixels of TOI-837 (i.e. those not accounted for by \texttt{PATHOS}) and applied a fixed correction term of $df = 0.069$ in the modelling. We imposed Gaussian priors on the projected rotational velocity ($v \sin {i_\star}$), the host star density ($\rho_\star$), and stellar radius ($R_\star$) from \citetalias{2024A&A...688A..15D}, while leaving the stellar rotation period ($P_{\rm rot}$) free to vary. We treated the stellar equatorial velocity ($v_{\rm eq}$, obtained from $P_{\rm rot}$) and inclination ($i_\star$) as independent variables. Therefore, we are not subject to the bias described in \citet{2020AJ....159...81M}. The $v \sin {i_\star}$ is a derived parameter. It is calculated from $v_{\rm eq}$ and $i_\star$ parameters at each step, and then compared with the prior. To explore the parameter space efficiently, we restricted $\cos i_\star$ to the range $[0, 1]$, effectively selecting a single stellar pole orientation. We treated the stellar limb darkening (LD) contribution by estimating $\rm u_1$ and $\rm u_2$ using \texttt{PyLDTk}\footnote{\url{https://github.com/hpparvi/ldtk}} \citep{2013A&A...553A...6H, 2015MNRAS.453.3821P}, taking into account the spectroscopic $T_{\rm eff}$ and $\log g$ from \citetalias{2024A&A...688A..15D}, assuming a boxcar filter as the passband in the ESPRESSO spectral range and adding $0.1$ in quadrature to their Gaussian errors to take into account the known model underestimation. Additionally, we used a linear law as a function of limb angle to model and fit the effect of stellar convective blueshift (CB) on the in-transit RV curve \citep[e.g.,][]{2016A&A...588A.127C}. Short-term stellar activity was incorporated into the model using a jitter term, which was added in quadrature to the RV and photometry errors. An extended description of how priors and boundaries have been defined can be found in \cite{2026arXiv260419179M}.

Our modelling of the RV anomaly foresees the calculation of a synthetic CCF, assuming a Gaussian profile for the local spectral line emerging from the surface of the star, and accounting for the missing contribution from the stellar surface covered by the planet. The model reproduces a CCF with characteristics similar to the observed one (such as the RV range and step), and fits a Gaussian to derive the predicted RV anomaly. This method has been introduced in \cite{2013A&A...554A..28C} and implemented in \texttt{PyORBIT} in \cite{2024A&A...684L..17M}. To minimise the differences between the RV measurement techniques and our modelling approach, we decided to analyse the RV measured from the CCF.  We note that the rotational profile differs from a Gaussian; we account for this bias by using a Gaussian profile on both the observed and modelled CCFs, thereby introducing the same systematic error on both sides.

We modelled stellar activity in the full 27.4 days \textit{TESS} photometric time series using a unidimensional GP. We adopted a rotation kernel as defined in \citet{celerite1, celerite2}. In contrast, due to the shorter (5-hour) time frame, we modelled the upward trend due to stellar activity observed in the spectroscopic time series using a second-degree polynomial function. We preferred a second-order over a linear one because the posterior of the quadratic coefficient is not centred on zero, though not significantly.

We performed a global optimisation of the parameters by running \texttt{PyDE} \citep{1997JGOpt..11..341S, 2016zndo.....45602P} for 100~000 generations and a Bayesian analysis of the RM signal in the RV time series using \textsc{emcee} \cite{2013PASP..125..306F} for 150~000 steps. The motivation behind the choice of using MCMC samplers is described in depth in \cite{2024A&A...684L..17M}. We used $4\times n_{\rm dim}$ walkers, with $n_{\rm dim}$ being the model dimensionality, and discarded the first 50~000 steps (burn-in). The Gelman-Rubin statistics \citep[][threshold value $\hat{R} = 1.01$]{1992StaSc...7..457G} were used to check the convergence of the chains. The best fitting values are given in Table \ref{table:model-rm}.

\subsection{Three-dimensional orbital obliquity}
\label{sec:psi}
\begin{figure}
   \centering
   \includegraphics[width=0.9\hsize]%
   {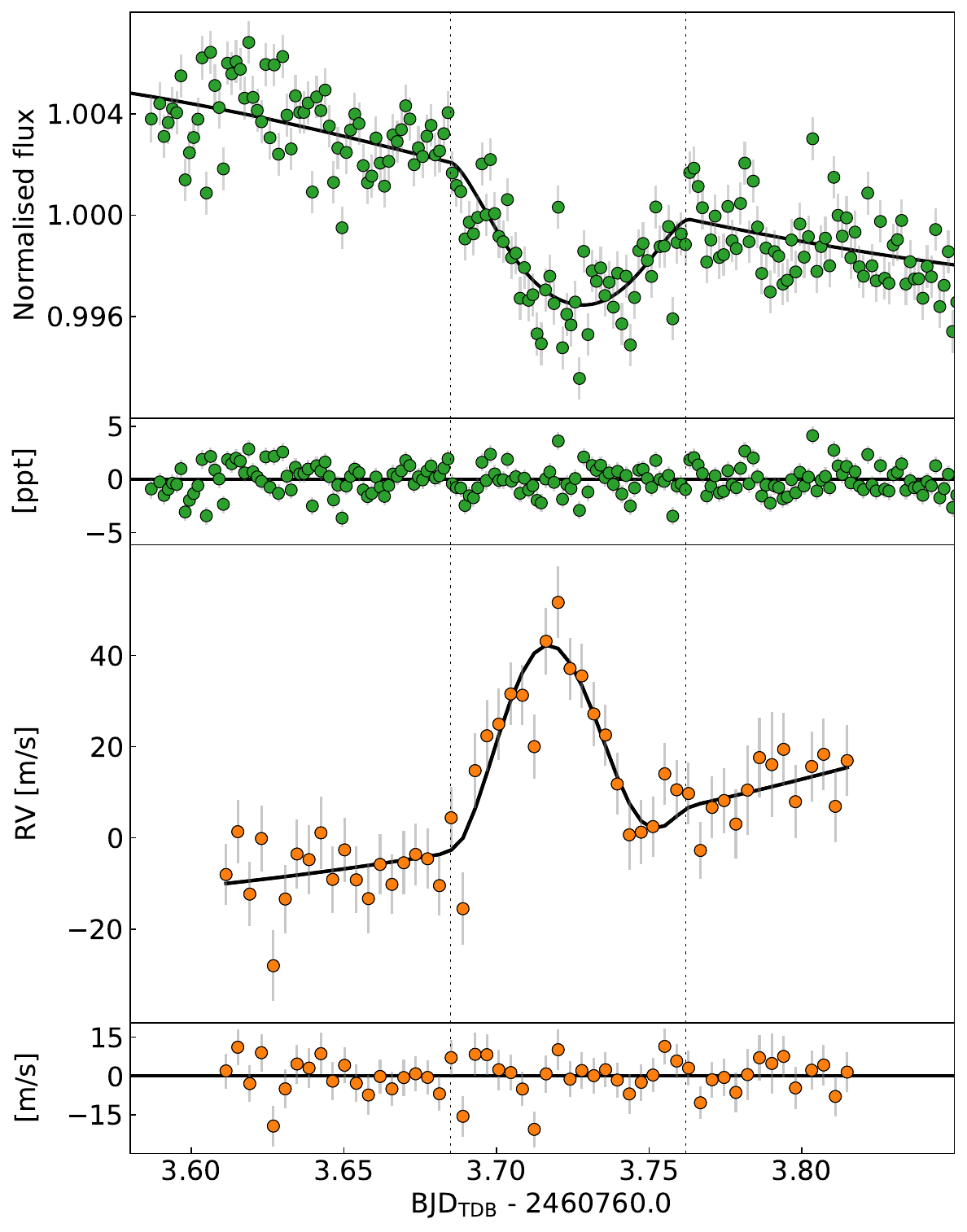}
   \caption{Simultaneous photometric and spectroscopic transit model fit of TOI-837~b, observed on 29 March 2025. The top two panels show the best-fit transit model (black line) overlaid on \textit{TESS} observations (green dots), and the residuals of the fit in ppt. The bottom panels instead show the best-fit RM model (black line) superimposed on the ESPRESSO RVs (orange dots), and the residuals from the fit in m s$^{-1}$. The RV data has been corrected for the systemic RV, while the upward trend caused by stellar activity has been modelled using a 2nd-order polynomial.}
   \label{fig:RM}
\end{figure}

By jointly modelling the space-based photometry and RM effect, we obtain a sky-projected obliquity $\lambda\, = 341.1^{+2.3}_{-2.5}$~deg and a reliable stellar rotation period $P_{\rm rot}\, = 3.00 \pm 0.02$~d, the latter being consistent with the values reported in \cite{2020AJ....160..239B} and \citetalias{2024A&A...688A..15D}. We then accurately translated $\lambda$ into the true three-dimensional orbital obliquity ($\psi$). We did so by sampling from the posterior distributions of the projected obliquity ($\lambda$), stellar inclination ($i_\star$) and planetary orbital inclination ($i_{\rm p}$), using Eq. (7) from \citet{2007AJ....133.1828W}, to estimate the three-dimensional orbital obliquity $\psi = 25.9 ^{+7.5}_{-6.3}$ deg. This result confirms the prograde motion of the planet and suggests a mildly misaligned orbit (Fig. \ref{fig:RM}).

\section{Discussion}

\subsection{Tidal decay of obliquity timescale}
The equilibrium tide model of \cite{2012ApJ...757...18A} gives an obliquity decay timescale exceeding the Hubble time for TOI-837. Such a result is confirmed also by considering the dynamical obliquity tides excited by the planet inside the star (see Appendix~\ref{tides_and_precession}). Therefore, the measured obliquity should still be unaltered by tidal effects, resulting in a direct diagnostic for the formation path of this planetary system.

\subsection{The mildly misaligned orbit of TOI-837~b}
\begin{figure}
   \centering
   \includegraphics[width=0.9\hsize]%
   {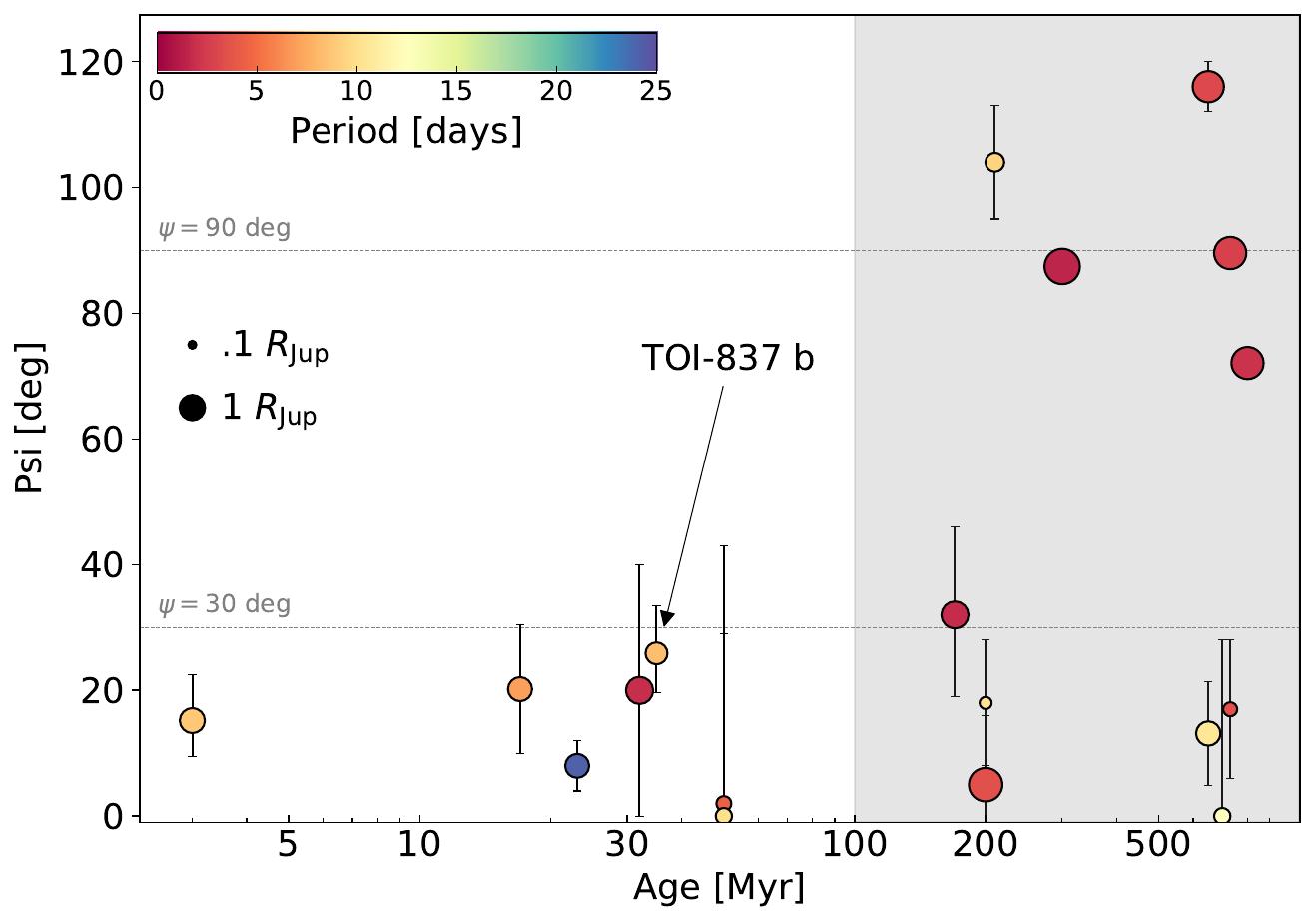}
   \caption{True obliquity for the youngest stars ($<$ 1 Gyr). The planetary orbital period is colour-coded, while the size of each dot indicates the radius of the given planet.}
   \label{fig:psi_age}
\end{figure}

Our analysis clearly shows that the orbit of the Saturn-size planet TOI-837~b is mildly misaligned. To contextualise this result, we selected all planets with a measured three-dimensional obliquity $\psi$ that orbit stars younger than 1 Gyr (see Fig. \ref{fig:psi_age}). Moreover, we divided the sample into young ($<$ 1 Gyr) and infant ($<$ 100 Myr) planets, as we are interested in investigating whether obliquity excitation for giant planets could occur during the early stages of a planetary system. TOI-837 b is the first planet younger than 100 Myr with accessible $\psi$ with a statistically significant, mildly misaligned orbit ($\approx 30$ deg). The mild but statistically significant misalignment of TOI-837~b, together with its very young age of around 35 Myr, supports the possibility of obliquity excitation taking place in the early stages of a planetary system. The obliquity of TOI-837 b has an extremely long tidal damping timescale, and it produces a precession of its orbital plane that should be measurable (see App. \ref{tides_and_precession} for details).

\subsection{Bound binary companion and dynamical implications}
TOI-837 has a co-moving stellar companion that is 4.92 mag fainter (in G band) and located at a separation of 2.3 arcseconds. \citetalias{2024A&A...688A..15D} confirmed that the co-moving star is bound (TOI-837~B, binary separation about 328 au). The gravitational pull of TOI-837~B might excite a high orbital inclination, consequently altering the obliquity through tidal and dynamical effects, and significantly influence the system's evolution. The accuracy and precision of our observations are unaffected, as the companion is too faint and distant to contaminate the spectra of TOI-837. Moreover, the RV variations are negligible in the time span of our observations.

The presence of a bound stellar companion helps us to place a few constraints on the formation and dynamical history of TOI-837~b. First, we inspected post-formation mechanisms involving a close stellar companion, capable of explaining mild misalignments. In particular, we estimated the timescale that stellar Kozai-Lidov oscillations \citep{2003ApJ...589..605W} would take if triggered by TOI-837~B. Taking into account the stellar mass of the companion ($M_{\rm B} = 0.39 M_\odot$) as reported in \citetalias{2024MNRAS.531.4275B}, we estimated a timescale of around 320 Myr \citep{2000ApJ...535..385F}, which is an order of magnitude longer than the age of the system. This result and the much faster relativistic precession of the apsidal line of the orbit make the stellar Kozai scenario improbable (see App. \ref{tides_and_precession}).

Conversely, a primordial misalignment scenario, such as that proposed by \cite{batygin2012}, could explain the moderate misalignment. In this framework, a massive distant companion exerts a secular torque on the protoplanetary disk, causing it to precess and become tilted with respect to the stellar spin axis, which is assumed to remain approximately fixed. Planets forming within the disk inherit this misalignment. This picture was later refined by \cite{picogna2015}, who showed that, once the planet becomes sufficiently massive, it can decouple dynamically from the disk due to the perturbations of the inclined companion. The planet and the disk then evolve almost independently, developing a significant mutual inclination. Nevertheless, inward migration is not halted: the planet continues to drift toward the star because its inclined orbit repeatedly crosses the disk, where interaction with the gas leads to angular momentum loss through dynamical friction.
An alternative scenario has been proposed by \cite{2022PASP..134h2001A}, in which the misalignment arises from a resonance crossing driven by a wide-orbiting, highly inclined stellar companion. Recent observations show that about one-third of isolated systems have misaligned outer discs (>10 au) from birth \citep{2025Natur.644..356B}, which challenges certain primordial obliquity excitation mechanisms. Nevertheless, the inner regions may remain aligned \citep[e.g.][and references therein]{2026arXiv260218553E}, and formation simulations by \citetalias{2024A&A...688A..15D} demonstrate that TOI-837 b formed in the inner disc between 2 and 4 au.

If such a misalignment is indeed primordial, then it is likely that TOI-837~b underwent disc-driven migration rather than HEM. This is further supported by the circular orbit of TOI-837~b and a circularisation timescale greater than 1 Gyr \citep{2006ApJ...649.1004A, 2008ApJ...678.1396J, 2010A&A...516A..64L}, which once again supports a primordial circular orbit \citepalias[in agreement with ][conclusions]{2024A&A...688A..15D}. This result is in line with the discussion first presented in \cite{2022PASP..134h2001A}, but it is an important addition given the statistically significant mild misalignment. Nevertheless, given the long eccentricity damping timescale, the orbit of TOI-837~b could have a small eccentricity that could be primordial or, in principle, be excited by the fast rotation of the star according to \cite{2010A&A...516A..64L}'s model because $(P_{\rm b}/P_{\rm rot}) \cos \psi > 18/11$. However, we show in App. \ref{tides_and_precession} that the excitation timescale is too long to be relevant, leaving any primordial eccentricity virtually unaffected.

%%%%%%%%%%%%%%%%%%%%%%%%%%%%%%%%%%%%%%%%%%%%%%%%%%%%%%%%%%%%%%
\section{Key findings}
\begin{enumerate}
    \item TOI-837~b is an infant giant planet (35 Myr) in the southern open cluster IC 2602 on a mildly misaligned orbit, and it is the only one known with such a configuration among infant (age < 100 Myr) planets with accessible $\psi$;
    \item Both the obliquity and circular orbit of TOI-837~b are unaltered by tidal effects, supporting the hypothesis that obliquity excitation occurred early-on in the formation and evolution of the system;
    \item A primordial misalignment scenario involving the bound stellar companion TOI-837~B, such as the exertion of a secular torque on the protoplanetary disc or resonance crossing, could explain the mild misalignment observed. This supports that TOI-837~b underwent disc-driven migration;
    \item Our transmission spectroscopy analysis yields no statistically significant absorption features nor evidence for detectable species in the planetary atmosphere.
\end{enumerate}

%%%%%%%%%%%%%%%%%%%%%%%%%%%%%%%%%%%%%%%%%%%%%%%%%%%%%%%%%%%%%%
\begin{acknowledgements}
Based on observations made with ESO Telescopes at the La Silla Paranal Observatory under programme ID P114.27GY. This paper includes data collected with the TESS mission, obtained from the MAST archive at the Space Telescope Science Institute (STScI). Funding for the TESS mission is provided by the NASA Explorer Program. STScI is operated by the Association of Universities for Research in Astronomy, Inc., under NASA contract NAS 5–26555. G.~M.~acknowledges support by the Space It Up project funded by the Italian Space Agency, ASI, and the Ministry of University and Research, MUR, under contract n. 2024-5-E.0 - CUP n. I53D24000060005.
\end{acknowledgements}

\bibliographystyle{aa} % style aa.bst
\bibliography{references} % your references 

%%%%%%%%%%%%%%%%%%%%%%%%%%%%%%%%%%%%%%%%%%%%%%%%%%%%%%%%%%%%%%%
% Appendices must be placed after   \end{thebibliography}
% They will be placed automatically on a new page.
%%%%%%%%%%%%%%%%%%%%%%%%%%%%%%%%%%%%%%%%%%%%%%%%%%%%%%%%%%%%%%%
\begin{appendix}
\nolinenumbers
%%%%%%%%%%%%%%%%%%%%%%%%%%%%%%%%%%%%%%%%%%%%%%%%%%%%%%%%%%%%%%%

\section{Probing the atmospheric composition}
\label{app:atmo}
\begin{figure*}
    \centering
    %\sidecaption
    \includegraphics[width=\textwidth]{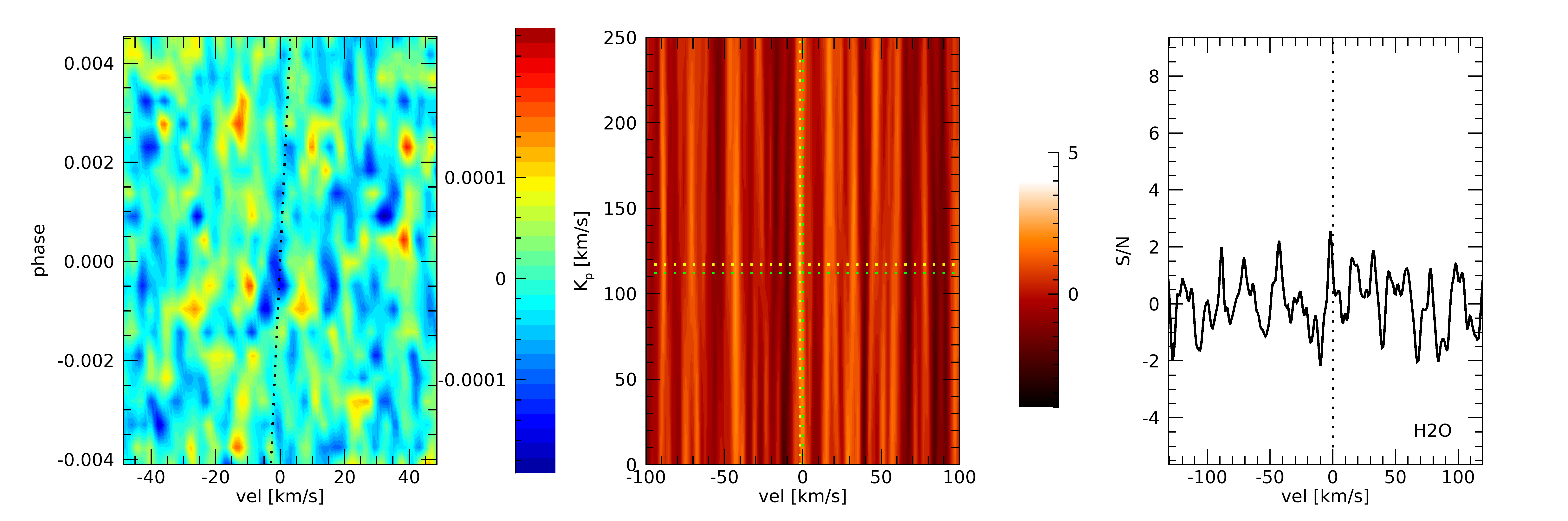}
    \caption{CCF signal for H$_2$O in TOI-837~b. \textit{Left panel:} 2D tomography of the in-transit CCF residuals. \textit{Central panel:} Kp-Vel map. The green dotted lines mark the theoretical planetary position. \textit{Right panel:} planetary absorption signal averaged at the theoretical Kp.}
    \label{fig:h2o}
\end{figure*}

Given its large transmission spectroscopy metric (TSM) value (see \cite{2018PASP..130k4401K}; \citetalias{2024A&A...688A..15D}), we tried to look at any possible atmospheric signal of TOI-837~b detectable in our data. We extracted the planetary transmission spectrum following the classic procedure used in high-resolution transmission spectroscopy \citep[e.g.][]{Wyttenbach2015,Borsa2021}. Telluric correction was already performed at DRS level, since for our analysis we used the "S1D$\_$TELL$\_$CORR" products of the pipeline. We shifted the spectra to the stellar restframe by using the Keplerian model of the system with the values in Table \ref{table:model-rm}, then we normalised each spectrum. Then we divided all the spectra by a master stellar spectrum created by averaging the out-of-transit spectra. At the end we moved to the planetary restframe by shifting all the residual spectra for the theoretical planetary radial velocity, and the transmission spectrum was then created averaging all the full-in-transit residual spectra. We searched for possible absorption given by Na D doublet, H$\alpha$, K, Mg, Li, without finding any trace of significant absorption beyond noise. We also tried to look for elements/molecules by using the cross-correlation technique. We generated model templates for V, Cr, Mg, K, Ca, H$_2$O by using \texttt{petitRADTRANS} \citep{pRT}, assuming an isothermal atmospheric profile with $T=1000$ K, continuum pressure level of 10 mbar and solar abundances. The atmospheric models were then translated into flux (R$_{\rm p}$/R$_{\rm s}$)$^2$, convolved at the ESPRESSO resolving power and continuum normalised. Cross-correlation between model templates and the single transmission spectra was then performed in the same way as in \citet{Borsa2021}. We find no evidence in the planetary atmosphere for the presence of any of the species we looked for. It is interesting to note the map relative to the H$_2$O non-detection (Fig. \ref{fig:h2o}). No spurious signal is present, a sign that the automatic telluric correction done by the ESPRESSO DRS is performing very well.

\section{Extra figures and tables}
\begin{figure*}[h]
   \centering
   \includegraphics[width=\textwidth]%
   {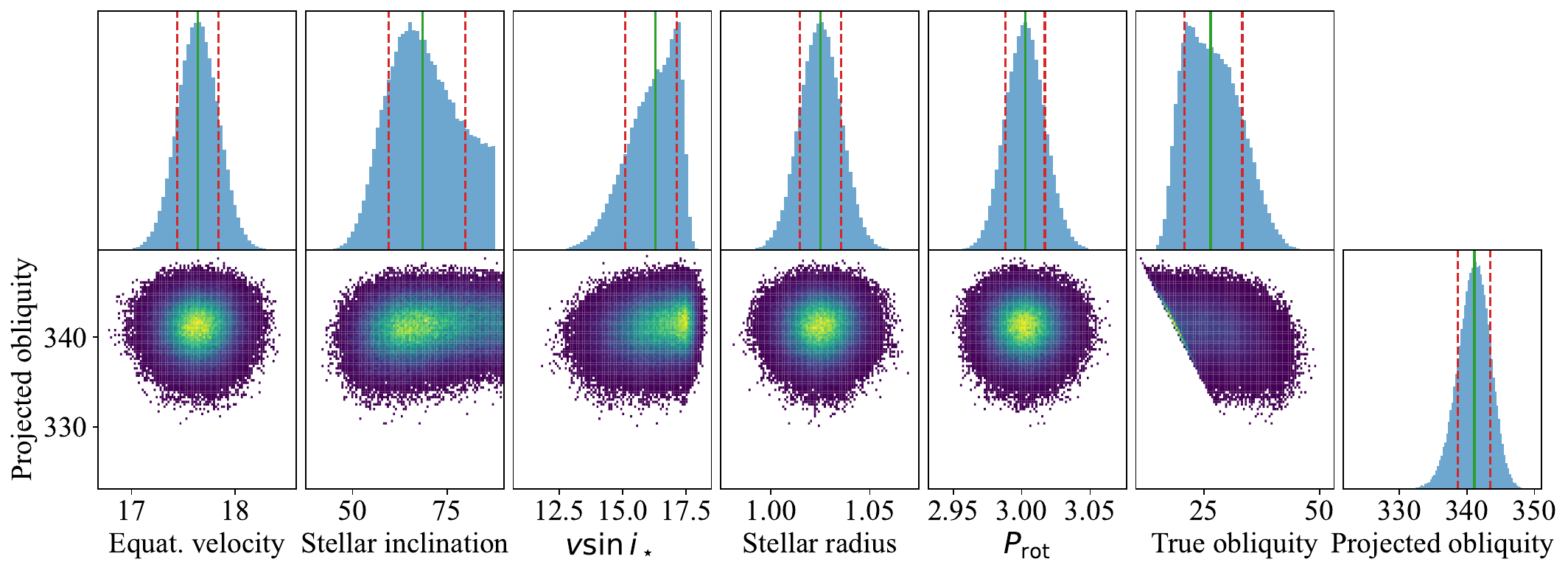}
   \caption{Posterior distributions of stellar parameters and obliquities that were retrieved in this work.}
   \label{fig:post}
\end{figure*}

This section includes Table \ref{table:model-rm} presenting all the priors and outcomes of the RM modelling and Figure \ref{fig:post} showing the retrieved posterior distributions.

\begin{table}[h]
\caption{Priors and outcomes of RM modelling.}             
\label{table:model-rm}    
\addtolength{\tabcolsep}{-0.5em}
\centering          
\begin{tabular}{l c c c} 
\hline\hline     
 
Parameter & Unit & Prior & Value \rule{0pt}{2.3ex} \rule[-1ex]{0pt}{0pt}\\ 
\hline 
\multicolumn{4}{c}{Planetary parameters} \rule{0pt}{2.3ex} \rule[-1ex]{0pt}{0pt}\\ 
\hline
   $P_{\rm b}$ & days & $\mathcal{U}$(8.32, 8.33) & 8.3249059(24) \rule{0pt}{2.3ex} \rule[-1ex]{0pt}{0pt}\\
   $T_{\rm 0,b}$ & BTJD\tablefootmark{a} & $\mathcal{U}$(2290.1, 2290.3) & 2290.21598(24) \rule{0pt}{2.3ex} \rule[-1ex]{0pt}{0pt}\\
   $a_{\rm b}/R_{\star}$ &  & ... & 17.62$\pm$0.14 \rule{0pt}{2.3ex} \rule[-1ex]{0pt}{0pt}\\
   $a_{\rm b}$ & au & ... & 0.0840$\pm$0.0009 \rule{0pt}{2.3ex} \rule[-1ex]{0pt}{0pt}\\
   $i_{\rm b}$ & deg & ... & 86.90$\pm$0.04 \rule{0pt}{2.3ex} \rule[-1ex]{0pt}{0pt}\\
   $e_{\rm b}$ &  & 0.0 & 0.0 
   \rule{0pt}{2.3ex} \rule[-1ex]{0pt}{0pt}\\
   $R_{\rm p}/R_{\star}$ &  & $\mathcal{U}$(0.0, 0.5) & 0.083$^{+0.007}_{-0.005}$ \rule{0pt}{2.3ex} \rule[-1ex]{0pt}{0pt}\\
   $b$ &  & $\mathcal{U}$(0.0, 0.99) & 0.952$^{+0.007}_{-0.006}$ \rule{0pt}{2.3ex} \rule[-1ex]{0pt}{0pt}\\
   $T_{14}$ & days & ... & 0.078$\pm$0.001 \rule{0pt}{2.3ex} \rule[-1ex]{0pt}{0pt}\\
   $R_b$ & $R_\oplus$ & ... & 9.30$^{+0.74}_{-0.56}$ \rule{0pt}{2.3ex} \rule[-1ex]{0pt}{0pt}\\
   $\lambda$ & deg & $\mathcal{U}$(0.0, 360) & 341.1$^{+2.3}_{-2.5}$ \rule{0pt}{2.3ex}  \rule[-1ex]{0pt}{0pt}\\
   $\psi$ & deg & ... & 25.9$^{+7.5}_{-6.3}$ \rule{0pt}{2.3ex} \rule[-1ex]{0pt}{0pt}\\
\hline

\multicolumn{4}{c}{Stellar parameters} \rule{0pt}{2.3ex} \rule[-1ex]{0pt}{0pt}\\ 
\hline 
   $\rho_{\star}$ & $\rho_{\odot}$ & $\mathcal{N}$(1.078, 0.028)\tablefootmark{b} & 1.061$\pm$0.026   \rule{0pt}{2.3ex} \rule[-1ex]{0pt}{0pt}\\
   $R_{\star}$ & $R_{\odot}$ & $\mathcal{N}$(1.036, 0.015)\tablefootmark{b} & 1.025$\pm$0.011   \rule{0pt}{2.3ex} \rule[-1ex]{0pt}{0pt}\\
   $P_{\rm rot}$ & days & $\mathcal{U}$(0.0, 5.0) & 3.002$\pm$0.016 \rule{0pt}{2.3ex} \rule[-1ex]{0pt}{0pt}\\
   $v \sin{i_\star}$ & km s$^{-1}$ & $\mathcal{N}$(16.5, 1.3)\tablefootmark{b} &  16.11$^{+0.92}_{-1.3}$   \rule{0pt}{2.3ex} \rule[-1ex]{0pt}{0pt}\\
   $u_1$ &  & $\mathcal{N}$(0.53, 0.10) & 0.62$\pm$0.09 \rule{0pt}{2.3ex} \rule[-1ex]{0pt}{0pt}\\
   $u_2$ &  & $\mathcal{N}$(0.15, 0.10) & 0.23$\pm$0.09 \rule{0pt}{2.3ex} \rule[-1ex]{0pt}{0pt}\\
   $\cos i_{\star}$ &  & $\mathcal{U}$(0, 1) & 0.36$^{+0.15}_{-0.21}$\rule{0pt}{2.3ex} \rule[-1ex]{0pt}{0pt}\\
   $i_{\star}$ & deg & ... & 68.9$^{+12}_{-9.8}$\rule{0pt}{2.3ex} \rule[-1ex]{0pt}{0pt}\\
   Conv. blueshift & & $\mathcal{U}$($-$2, 0)& $-$1.05$^{+0.70}_{-0.66}$\rule{0pt}{2.3ex} \rule[-1ex]{0pt}{0pt}\\
   $v_{\rm eq}$ & km s$^{-1}$ & $\mathcal{U}$(0, 70) & 17.28$\pm$0.21 \rule{0pt}{2.3ex} \rule[-1ex]{0pt}{0pt}\\
   \hline
   \multicolumn{4}{c}{Modelling parameters} \rule{0pt}{2.3ex} \rule[-1ex]{0pt}{0pt}\\ 
\hline
RM jitter & m s$^{-1}$ & ... & 1.8 $^{+1.6}_{-1.2}$ \rule{0pt}{2.3ex} \rule[-1ex]{0pt}{0pt}\\
RM offset & m s$^{-1}$ & ... & 17775.3 $^{+1.7}_{-1.5}$ \rule{0pt}{2.3ex} \rule[-1ex]{0pt}{0pt}\\
   \cline{0-1} 
   \textit{Polynomial trend} & & & \rule{0pt}{2.2ex} \rule[-0.8ex]{0pt}{0pt}\\
   Coefficient c$_1$ &  & ... & 125 $\pm$ 19 \rule{0pt}{2.3ex} \rule[-1ex]{0pt}{0pt}\\
   Coefficient c$_2$ &  & ... & 254 $^{+173}_{-273}$ \rule{0pt}{2.3ex} \rule[-1ex]{0pt}{0pt}\\
\hline   
\end{tabular}
\tablefoot{
\tablefoottext{a}{BTJD = BJD$_{\rm TDB}$ - 2457000.0.}
\tablefoottext{b}{\citet{2024A&A...688A..15D}.}
}
\end{table}

\begin{figure}[h]
   \centering
   \includegraphics[width=\columnwidth]%
   {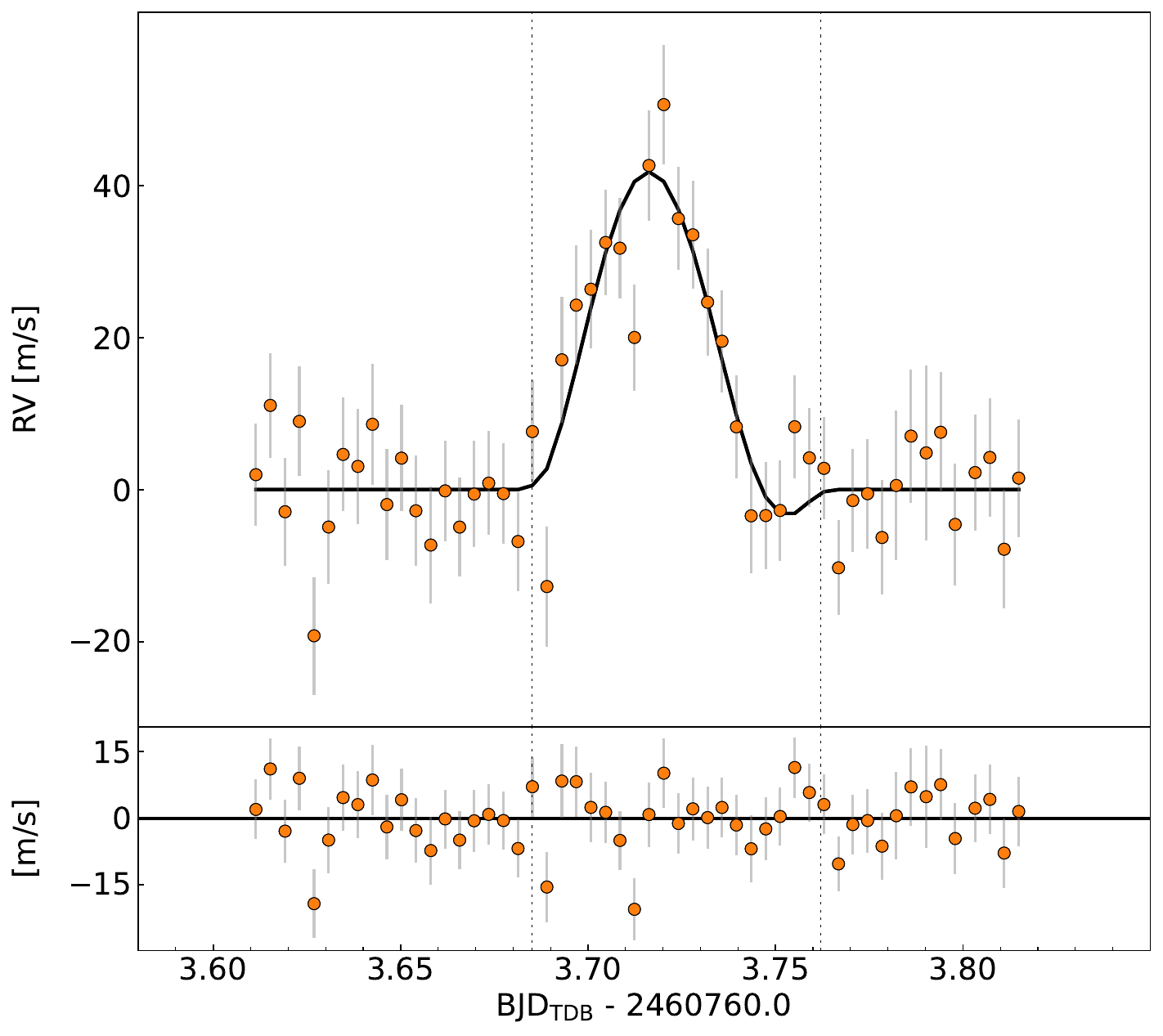}
   \caption{Figure similar to Fig. \ref{fig:RM} but with the RV data corrected for both systemic RV and stellar activity.}
   \label{fig:rm_trend}
\end{figure}

\section{Tides and precession}
\label{tides_and_precession}
The obliquity of the star in our system can be effectively damped by dynamical obliquity tides according to the mechanism proposed by \citet{Lai12}. Considering a stellar modified tidal quality factor $Q^{\prime}_{\rm s} = 10^{6}$, that corresponds to an efficient tidal dissipation inside the star itself, we obtain an e-folding decay time of the stellar obliquity of $\sim 9 \times 10^{12}$~years, much longer than the Hubble time. To compute such a timescale, we used the constant tidal time lag model by \citet{2010A&A...516A..64L} translating the modified tidal quality factor into a time lag with the relationship $\Delta t = (3/2)(k_{\rm L2} n Q^{\prime}_{\rm s})^{-1}$, where $k_{\rm L2}$ is the stellar potential Love number of degree two and $n$ is the orbital mean motion. 

TOI-837~b is one of the few giant close-in planets for which the orbital period is longer than the rotation period of the host star, making tides transfer angular momentum from the stellar spin to the orbit of the planet. However, such a transfer is very slow because of the relatively large value of $a/R_{\rm s} = 17.62 \pm 0.14$. Therefore, the timescales for the increase of the semimajor axis and for the excitation of the eccentricity of the orbit are remarkably longer than the Hubble time, even considering dynamical tides inside the star. Dynamical tides in the form of inertial waves, having the Coriolis force as their restoring force, are expected to be excited inside the star because the tidal frequency $2 |n-\Omega_{\rm s}| < 2\Omega_{\rm s}$, where $\Omega_{\rm s}$ is the stellar spin frequency \citep[e.g.,][]{2014ARA&A..52..171O}. 

The tidal damping of the orbital eccentricity is dominated by the dissipation of tides inside the planet. Considering a modified tidal quality factor of the planet $Q^{\prime}_{\rm p} =10^{5}$, that is similar to that of Jupiter, we obtain an eccentricity damping timescale  $\tau_{\rm e} \equiv e/|de/dt| = 3.4$~Gyr indicating that any primordial small eccentricity could have not been appreciably damped along the lifetime of the system ($\sim 35$~Myr). On the other hand, the timescale for the synchronization of the planetary rotation with the orbital period and the damping of any primordial obliquity of the planetary spin is only $\sim 0.14$~Myr, much shorter than the age of the system. 

The fast rotation of TOI-837 and the obliquity of the orbit of its planet produce a precession of the line of the nodes of its orbit that can be evaluated using, for example, the model by \citet{DamianiLanza11}. Using their Eq.~(8), we find a period of the precession of the line of the nodes of $0.85$ or $1.7$~Myr for a stellar apsidal motion constant $k_{2}=0.02$  or $k_{2}=0.01$, respectively. According to the models computed by \citet{Claret19}, the former value of the apsidal motion constant is close to that of the present Sun, while the latter is for a star on the zero-age main sequence,  $\sim 20$\% more massive than the Sun and with a solar chemical composition. 

The precession of the orbital plane of TOI-837~b causes a change in the inclination of its orbit to the line of sight of $\sim 0.''69$/yr for $k_{2} =0.01$. This produces a change in the duration of the transit of $\sim 14$~s in ten years. Detecting such a change might in principle be possible with the CHEOPS space-craft, where e.g. \cite{2021MNRAS.506.3810B} obtained a precision of 13-16 s, although we note that this was for a bright, less active system.

The period of precession of the spin of the planet is only $\sim 67$ years, considering synchronization with the orbital motion, and a planetary Love number and normalized moment of inertia equal to those of Jupiter. This implies that the planetary spin precession cannot be captured into a resonance with the nodal precession during the evolution of the system \citep{MillhollandLaughlin19}. Therefore, the planetary spin is predicted to stay normal to the orbital plane during all the future evolution of the system and a large planetary obliquity is not expected to be excited by such a mechanism. 

If the orbit of TOI-837~b has a small eccentricity, say of 0.05, its apsidal line will precede under the action of the general relativistic effects and the stellar quadrupole moment. We estimate an apsidal precession period of $\sim 45 400$ years with more than 80\% of the precession rate produced by general relativity effects \citep[e.g.][]{MardlingLin02}. Such a relativistic precession does not affect the line of the nodes that precedes under the effect of the stellar gravitational quadrupole moment only ($J_{2} \sim 10^{-5}$ for $k_{2} = 0.01$). As a consequence of the apsidal precession, the mid transit times of the planetary transits and occultations oscillate along the precession cycle \citep[e.g.][]{DamianiLanza11}. However, the effect is very small and can probably not be detected because the transit time variation in ten years is only 0.05~s for an eccentricity of 0.05. By considering an eccentricity of 0.1, the transit time variation doubles to $\sim 0.10$~s, while all the precession periods and the transit duration variation remain the same within a few percent. 

We conclude that the apsidal precession is always much faster than the Kozai cycles potentially induced by the distant stellar companion TOI-837~B, which implies that any excitation of the eccentricity of TOI-837~b by the gravitational perturbation of that companion with a Kozai mechanism is negligible. The eccentricity excited by TOI-837~B along the main-sequence lifetime of the system is also negligible in the case of nearly coplanar orbits, according to the model by \citet{Mardling07}, because of the large separation between the planet and the distant companion.

\end{appendix}
\end{document}